\documentclass[aps,prb,twocolumn,floatfix,footinbib,showpacs,superscriptaddress]{revtex4-1}
\usepackage{graphicx}
\usepackage{amsfonts,amsmath,amssymb}
\usepackage{amsthm}
\usepackage{dsfont,bm}
\usepackage{color}
\usepackage{soul} 
\usepackage{amsbsy}
\usepackage[colorlinks=true,linkcolor=blue,pagecolor=blue,filecolor=blue,menucolor=blue,urlcolor=blue,citecolor=blue,anchorcolor=blue]{hyperref}

\usepackage{mathbbol}

\usepackage{sidecap}

\begin{document}

\title{Multiscale statistical quantum transport in porous media and random alloys with vacancies}

\author{Elham Sharafedini}
\email{elsharafedini@email.kntu.ac.ir }
\affiliation{Department of Physics, K.N. Toosi University of Technology, Tehran 15875-4416, Iran}
\author{Hossein Hamzehpour}
\email{hamzehpour@kntu.ac.ir }
\affiliation{Department of Physics, K.N. Toosi University of Technology, Tehran 15875-4416, Iran}
\author{Mohammad Alidoust}
\email{phymalidoust@gmail.com}
\affiliation{Svalevegen 5, 7022, Trondheim, Norway}
\affiliation{Department of Physics, K.N. Toosi University of Technology, Tehran 15875-4416, Iran}
\date{\today}

\begin{abstract}
We have developed a multi-scale self-consistent method to study the charge conductivity of a porous system or a metallic matrix alloyed by randomly distributed nonmetallic grains and vacancies by incorporating Schr\"{o}dinger's equation and Poisson's equation. To account for the random distribution of the nonmetallic grains and clusters within the alloy system, we have used an uncorrelated white-noise Monte-Carlo sampling to generate numerous random alloys and statistically evaluate the charge conductance. We have performed a parametric study and investigated various electrical aspects of random porous and alloy systems as a function of the inherent parameters and density of the random grains. Our results find that the charge conductance within the low-voltage regime shows a highly nonlinear behavior against voltage variations in stark contrast to the high-voltage regime where the charge conductance is constant. The former finding is a direct consequence of the quantum scattering processes. The results reveal the threshold to the experimentally observable quantities, e.g., voltage difference, so that the charge current is activated for values larger than the threshold. The numerical study determines the threshold of one quantity as a function of the remaining quantities. Our method and results can serve to guide future experiments in designing circuital elements, involving this type of random alloy system.  

\end{abstract}
                             
\maketitle

\section{Introduction} \label{sec1}

Due to the high importance of the electrical transport properties of heterogeneous systems, the charge transport process in disordered media at different length scales has been an active area of research for decades, both experimentally and theoretically\cite{sahimi1,A.A.Snarskii1,M.Walschaers}. These materials platforms include amorphous semiconductors, ionic conductive glasses,
ionic or electronic conducting polymers, metal-cluster compounds, organic semiconductors, and nonstoichiometric or polycrystals \cite{M.Kim,S.Torquato,E.Pazhoohesh,E.N.Oskoee,E.Sharafedini,J.C.Dyre,M. Janssen}. 

The scale size highly influences the end properties of electronic devices. For example, within the nanoscale size ($\sim10^{-9}$~m) quantum mechanical phenomena, such as quantum entanglement, charge trapping, and quantum tunneling are dominating scenarios to study and predict the electronic and thermodynamic behavior of devices. When transitioning from the nanoscale size into the mesoscale size ($\sim10^{-6}$~m), the direct consequences of the quantum mechanical phenomena on the characteristics of electronic devices become weaker. 
\\*
One main reason is that, by nature, the strength of quantum mechanical phenomena diminishes with increasing the size of the system. Other major factors that increasingly become important with increasing the size of the system include the impact of the non-uniformity and spatial localization of compositions and clusters. This effect is specifically important when dealing with alloys made of different materials and compositions. Naturally, the compositions are randomly distributed across a parent matrix that should be properly accounted for in simulations in order to have reliable predictions. Oftentimes, to model and make predictions of the random alloys, effective models for the electron transport based on the drift-diffusion are employed \cite{J.C.Dyre,C.DeFalco1,S.Yamakawa,L.Shifren1}. Also, employing machine learning techniques might be helpful \cite{K.Li1,N.A.Lanzillo1,R.Korol1,J.G.Nedell1,A.K.Pimachev1}. However, these effective models are unable to account for the quantum mechanical details of an alloy system and basically deal with few effective parameters \cite{A.DiVito1}. For instance, one might consider these randomly distributed solid solutes and compositions (and, in general, any other disorder and impurities) as electron scatterers, simulate them by white-noise Gaussian potentials, and derive diffusion-like continuous models from an effective model Hamiltonian. In this case, the influence of the random scatterers is abstracted into a diffusion constant \cite{R.Landauer1,N.F.Mott1,G.Eilenberger1,K.D.Usadel1}. Although this approach was recently generalized and formulated for various materials platforms, such as spin-orbit coupled materials, topological insulators, black phosphorus, and Weyl semimetals, it can only describe limited situations properly where the electron scatterers can be simply modeled with nonmagnetic localized potentials  \cite{M.Alidoust1,M.Alidoust2,M.Alidoust3,M.Alidoust4,M.Alidoust5,M.Alidoust6,M.Alidoust7,M.Alidoust8}. 

One way to address the device characteristics highly accurately at the nanoscale size is the atomistic models. The atomistic models can to some extent allow for taking the quantum interactions of atoms and compositions, strain, and lattice mismatch into calculations \cite{A.K.Pimachev1,V.S.Proshchenko1,V.S.Proshchenko2,M.ODonovan1}. However, the drawback of this approach is twofold. First, the number of atoms that can be considered in simulations goes not far beyond a few hundreds, and, thus, the associated simulations remain within the nanoscale size. Second, it is a highly formidable task, if not impossible, to take strain tensor and lattice mismatch at interfaces between materials precisely aligned with the experiment into the calculations. Additionally, in the device form, the inclusion of millions of atoms, proper strain tensor, and lattice mismatch, and a sufficiently large number of random sampling make the simulations highly expensive and challenging. Accordingly, the atomistic scale calculations are often limited to ideal crystals that rarely are achievable in experiments. To overcome the length scale limitation and to be able to transit successfully into the mesoscale size, one may resort to multiscale approaches \cite{M.AufderMaur1,S.Steiger1,C.Yam1,M.AufderMaur2,A.David1,T.-Y.Tsai1,H.-H.Chen1,V.Vargiamidis1,M.ODonovan1}. These approaches usually operate collaboratively, namely, the results of a lower scale size calculation are used as inputs to a higher scale size formalism. Generally, the higher scale size formulations deal with less parameters than the lower scale size calculations, the atomistic details of devices are, thus, weakened at the cost of having larger scale calculations. 

In effect, there are several parameters that affect the electronic transport  properties of a random alloy. Among the most important parameters are the density of compositions, chemical potential, and the thickness of compositional grains. Therefore, to achieve reliable designed heterostructures one needs to determine the complex links of these parameters to the electronic effective performance of devices.      
\\* 
In order to more accurately and realistically simulate the random alloy systems and account for the quantum mechanical effects, we have developed a self-consistent approach to obtain the spatial distribution of a driving electrostatic potential across a finite-sized sample and calculate the local charge conductance iteratively. We have incorporated a large number of statistically random samplings during the calculations and taken an average over all results during the calculations to obtain the effective charge conductivity. Considering a two-phase porous or random alloy system, mixing an insulator phase and a moderate bandgap compound, the results reveal a highly nonlinear charge conductance against the voltage difference, bandgap, the size of grains, and the density of the two phases. 
It is found that the charge conductance is only activated within a specific region of parameter space. By performing a parametric study, we determine the threshold borders of two representative observable quantities, i.e., the voltage difference and density of the two phases, against the other experimentally relevant parameters available in the system. The connection between the microscopic and macroscopic models in our self-consistent approach allows us to study systems with mesoscale sizes and yet to keep the quantum mechanical effects such as resonances and tunneling effects. Moreover, accounting for the capacitance effect, our approach is capable of providing a framework to study systems subject to an ac voltage with a given frequency and account for the spatial charge accumulation around the grains and clusters, which is relevant for the ac regime.

The paper is organized as follows. In Sec.~\ref{sec2}, we describe the developed multi-scale self-consistent approach in detail. In Sec.~\ref{results}, we present the results of the algorithm, and discuss various aspects of the charge conductance in a two-phase random porous and alloy system against differing experimentally relevant parameters available in the system. Additionally, we illustrate how one is able to obtain the functionality of the threshold value of an experimental parameter to the remaining parameters. Finally, we give concluding remarks in Sec.~\ref{conclusion}.

\section{Approach and formalism}\label{sec2}
We first introduce the macroscopic model used to obtain the governing equations to the dc charge conductivity in heterogeneous materials in Sec.~\ref{mac}. Next, in Sec.~\ref{mic}, we discuss the calculation of local charge conductivity through Schr\"{o}dinger's equation, providing quantum mechanical factors to our approach. In Sec.~\ref{algor}, the self-consistent algorithm is outlined. Finally, in Sec.~\ref{sampling}, the generation of the two-phase random alloys, how to account for the randomness in calculating the electrostatic potential, and the statistical evaluations are described.

\subsection{The macroscopic theory}\label{mac}

In the macroscopic model, it is assumed that the random alloy system hosts moving charged carriers that can be described by a spatially varying and frequency-independent local conductivity $g({\bf r})$. Accordingly, the constitutive equations of this model can be expressed by  

\begin{subequations}
\begin{gather}
\mathbf{J}(\mathbf{r},t)=-g(\mathbf{r},\varphi)\mathbf{\nabla}\varphi(\mathbf{r},t)\;,\label{eq1}\\
\mathbf{D}(\mathbf{r},t)=-\epsilon_\infty\mathbf{\nabla}\varphi(\mathbf{r},t)\;\label{eq2},
\end{gather}
\end{subequations}
where $\mathbf{J}(\mathbf{r},t)$, $\mathbf{D}(\mathbf{r},t)$, $\varphi(\mathbf{r},t)$, and $\epsilon_\infty$ are the charge current density, displacement vector, electrostatic
 potential, and the permittivity of the system within the high enough frequencies, respectively. 
These equations are accompanied by Gauss' law
\begin{equation}
\mathbf{\nabla \cdot D}(\mathbf{r},t)=\rho(\mathbf{r},t)\;,\label{eq3}
\end{equation}
and the charge continuity equation
\begin{align}
\dfrac{\partial\rho(\mathbf{r},t)}{\partial t} +\mathbf{\nabla\cdot J} (\mathbf{r},t)=0\;,\label{eq4}
\end{align}
in which $\rho(\mathbf{r},t)$ is the charge density available in the system. Combining these equations, one arrives at the following equation for the electrostatic potential:
\begin{align}
\mathbf{\nabla} \cdot [\epsilon_\infty\dfrac{\partial}{\partial t}\mathbf{\nabla} \varphi(\mathbf{r},t)+
g(\mathbf{r},\varphi)\mathbf{\nabla}\varphi(\mathbf{r},t)]=0\;.\label{eq5}
\end{align}
We consider a dc equilibrium situation where the continuity equation results in 
 \begin{equation}
 \mathbf{\nabla\cdot}{[g(\mathbf{r},\varphi)\mathbf{\nabla}\varphi(\mathbf{r})]}=0\;.\label{eq:poisson}
 \end{equation}
We use the finite volume method to solve Eq. (\ref{eq:poisson}). In the finite volume method, one first integrates Eq. (\ref{eq:poisson}) over the system volume. Next, invoking the divergence theorem, the volume integral is converted into a surface integral,
\begin{equation}
\begin{aligned}
\int_{V }\mathbf{\nabla} \cdot{[g(\mathbf{r},\varphi)\mathbf{\nabla}\varphi(\mathbf{r})]}dv\\
=\int_{\partial V}g(\mathbf{r},\varphi)\mathbf{\nabla}\varphi(\mathbf{r})\cdot d\mathbf{A}=0\;, \label{eq8}
\end{aligned}
\end{equation}
where $d\mathbf{A}$ and $\partial V$ are the surface element and the system surface, respectively. The second integral in Eq. (\ref{eq8}) implies that the net current passing through the entire system and each element is conserved. Therefore, accounting for the four sides of a two-dimensional rectangular control volume, one can write
\begin{equation}
\int_{\partial V}\mathbf{J} \cdot d\mathbf{A} \simeq \sum_{k=1}^4 J_k A_k =0\;,
\label{eq9}
\end{equation}
where {\bf J} is the total current flux on each interface.
In Eq. (\ref{eq9}), each lattice block is centered at $(i, j)$ with four directions in which $k=1-4$ denotes $\pm\hat{x}, \pm\hat{y}$, and, therefore, we arrive at the grid schematically shown in Fig.~\ref{schem}(a)\cite{sahimi1, J.C.Dyre}. The spatial dependency of the static potential $\phi({\bf r})$ is determined by the values of potential at the grid points (marked by a circle) through a self-consistent approach that shall be discussed in detail below. Hence, the lines connecting the lattice grids display the material filling between two grid points.

When applying an ac voltage to a system, the capacitance effect contributes more significantly to the charge conductivity. This effect originates from the charge accumulation around the grains and clusters. To account for the capacitance effect, one needs to renormalize the local charge conductance at each iteration and add the contribution of the frequency-dependent ac conductivity \cite{J.C.Dyre}. Nevertheless, in this paper, we study the dc conductivity and neglect the contribution of the capacitance effect.   

\subsection{The microscopic theory}\label{mic}

The moving particles at the microscopic level obey the associated low-energy Hamiltonian. Within the low-energy regime and conventional materials with parabolic band structures, the effective Hamiltonian reads 
\begin{equation}
H({\bf r}) = \frac{{\bf p}^2}{2m^*} + U({\bf r}).
\label{Hamil}
\end{equation}
The momentum of moving quasiparticles with effective mass $m^*$ is given by ${\bf p}=-i \hbar  (\partial_x,\partial_y,0)$
 and the chemical potential is denoted by $U({\bf r})$. In our calculation, the effective mass is considered to be location independent. Therefore, throughout the following calculations, only the chemical potential is considered to be location dependent when solving the local Schr\"{o}dinger equation (\ref{Hamil}) and obtaining the spatial maps of electrostatic potential self-consistently.  

To calculate the flow of charged moving particles through the grains and cluster regions, one should work in configuration space. Throughout the calculations performed in this paper, we consider two-dimensional random alloy systems, i.e., $\mathbf{ r}\equiv(x,y,0)$. Setting the quantum mechanical definition of the time variation of charge density to zero, i.e., $\partial_t\rho_\text{c}\equiv 0$, one finds the following expression in terms of the low-energy effective Hamiltonian,
\begin{equation}\label{crntdif}
\begin{split}
\frac{\partial \rho_\text{c}}{\partial t}=\lim\limits_{\mathbf{r}\rightarrow \mathbf{r}'}\frac{1}{i \hbar}\Big[ \psi^{*} (\mathbf{r}'){H}(\mathbf{r})\psi(\mathbf{r})-\psi^{*}(\mathbf{r}'){H}^\dag(\mathbf{r}')\psi(\mathbf{r})\Big].
\end{split}
\end{equation}
Assuming no charge sink or source and applying the current conservation law, the charge current
 at the microscopic level reads,
\begin{align}\label{Miccurrent}
{ J}_k A_{k}=I_k=\dfrac{i}{\hbar} \int \hspace{-.1cm} d\mathbf{r}\Big\{\psi^{*}(\mathbf{r}) \overrightarrow{{H}}(\mathbf{r}) \psi(\mathbf{r})-
\psi^{*}(\mathbf{r}) \overleftarrow{{ H}}(\mathbf{r})\psi(\mathbf{r}) \Big\}.
\end{align} 
Here, the Hamiltonian is given by Eq.~(\ref{Hamil}). In this notation, the arrow directions indicate the specific wavefunctions that the Hamiltonian acts on. 

\begin{figure}[b]
\begin{center}
\includegraphics[width=0.4\textwidth]{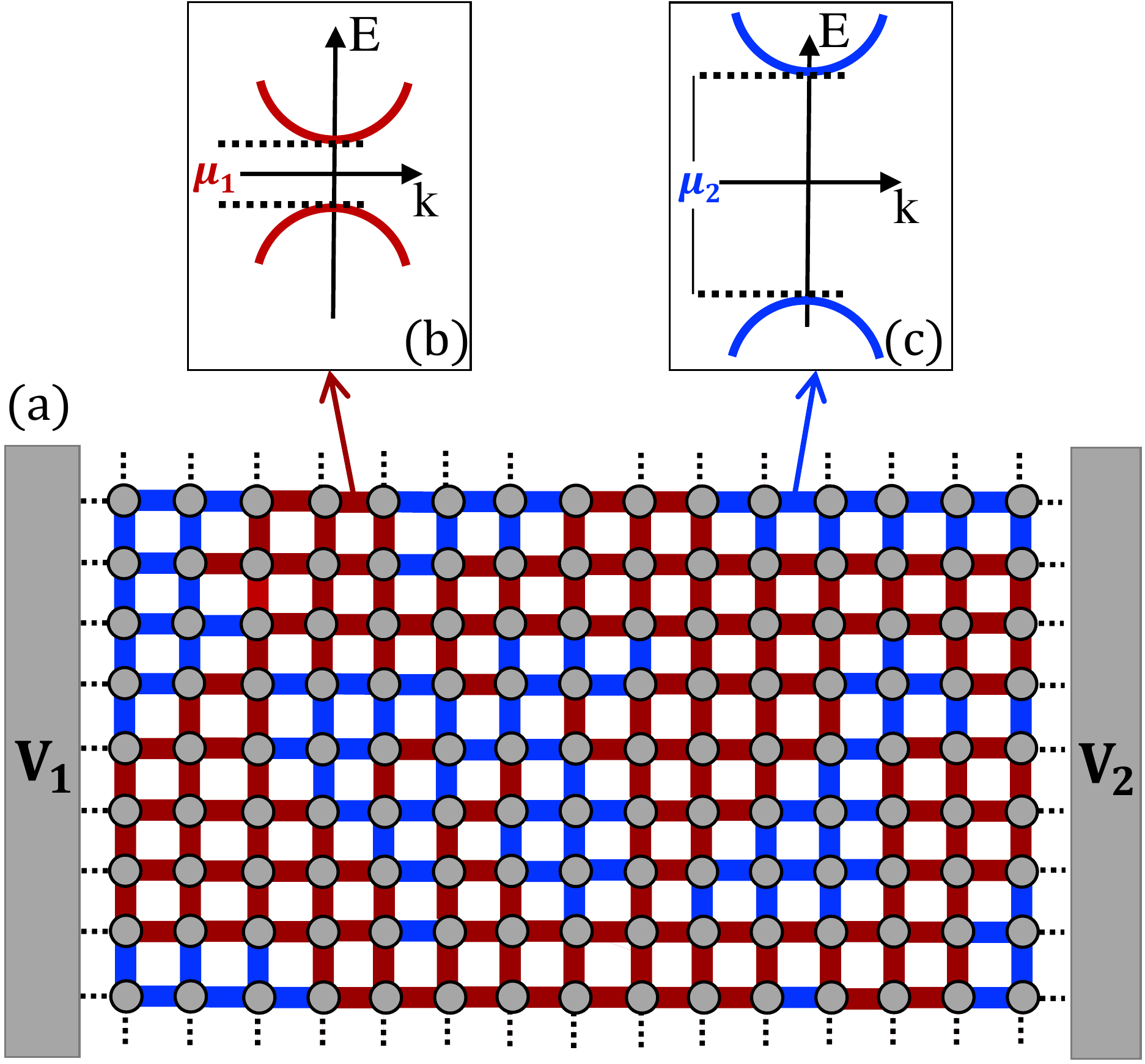}
\caption{(a) Schematic of the gridded random porous or alloy system considered. The system is made of two different phases marked by blue and red regarded as resistors or potential barriers with different  resistivities $g^{-1}(r)$. The two phases are randomly distributed across the sample. The entire sample is sandwiched between two electrodes with voltages V1 and V2. Circles/points are potential representatives of lattice blocks. (b) and (c) show the band structure of the blue and red regions. It is assumed that the bandgap of the blue regions is much larger than the bandgap of the red regions, i.e., $\mu_2\gg\mu_1$.}
\label{schem}
\end{center}
\end{figure}

In order to obtain the local charge current at each lattice link
, one should solve Schr\"{o}dinger's equation $H \psi({\bf r})=E \psi({\bf r})$ together with proper boundary conditions. The boundary conditions we use are the continuity of the wave functions and their first order derivative, ensuring the continuity of the velocity of the moving particles. Once Schr\"{o}dinger's equation is solved and the wavefunction $\psi({\bf r})$ is obtained, one inserts $\psi({\bf r})$ into Eq.~(\ref{Miccurrent}), determines the local charge current, and thereby the local conductance through ${I}_k ({\bf r})=g({\bf r}, \delta_{k} \varphi) \delta_k \varphi$, where $\delta_{k} \varphi =\varphi_{k}-\varphi_{k'}$ is the potential difference between two neighboring grid points in Fig.~\ref{schem}(a). Note that in the presence of finite temperature, one can take it into account by the BTK formalism $I_{k}({\bf r})= \int g({\bf r},\delta_{k} \varphi) [f(E-e \varphi{_k})-f(E-e \varphi_{k'})] dE$, where $f$ is the Fermi distribution function. We emphasize that nonzero temperature suppresses the conductivity as it does in any conventional system.\cite{sahimi1,J.C.Dyre}  In our calculations throughout the paper, we normalize the charge conductance by $\sigma_0 = e^2  h^{-1}$, which is the conductance unit of a single channel perfect conductor.

In this paper, we consider a two-phase random heterogeneous system sandwiched between two electrodes as shown in Fig.~\ref{schem}.
 The left and right electrodes are displayed by two gray regions and possess voltages $V_1$ and $V_2$, respectively, that allow us to define the voltage difference $V=V_2-V_1$. As shown in Figs.~\ref{schem}(c), it is assumed that the blue region is described by a larger bandgap $\mu_2$ than that of the red region $\mu_1$ that can prevent any charge current at low enough voltage differences. Hence, the blue region represents insulator grains or vacuum, simulating a porous system or alloy with vacancies. The red region has a moderate bandgap $\mu_1$ that depending on its magnitude $|\mu_1|$, the voltage difference across the sample $V$, and the thickness of the red grains, it can pass charged particles through the quantum transport process. Therefore, the red regions can simulate intermetallic solid solutes in a random alloy system. In what follows, we consider a situation where $\mu_2\gg \mu_1$ and for simplicity in our notation we define $\mu=\mu_1$.
\begin{figure}[t!!]
\includegraphics[width=7cm]{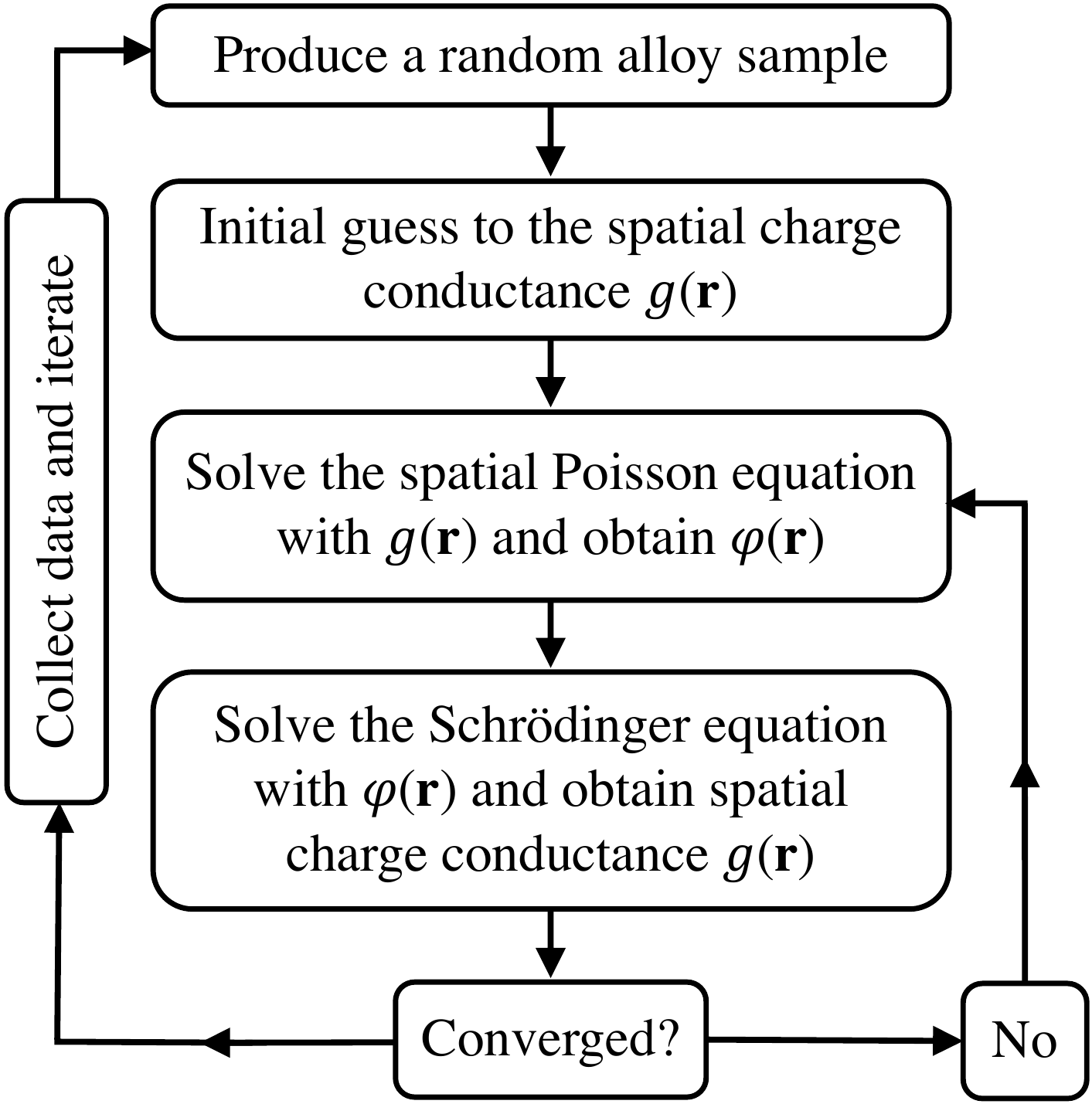}
\caption{The flow chart diagram of the self-consistent approach to obtain the spatial profile of charge conductance $g(\mathbf{r})$ and electrostatic potential $\varphi(\mathbf{r})$ in a random alloy system with a given set of the density of various cluster compounds available in the sample. }
\label{flowchart}
\end{figure}

\begin{figure*}[htpb]
\includegraphics[width=\textwidth]{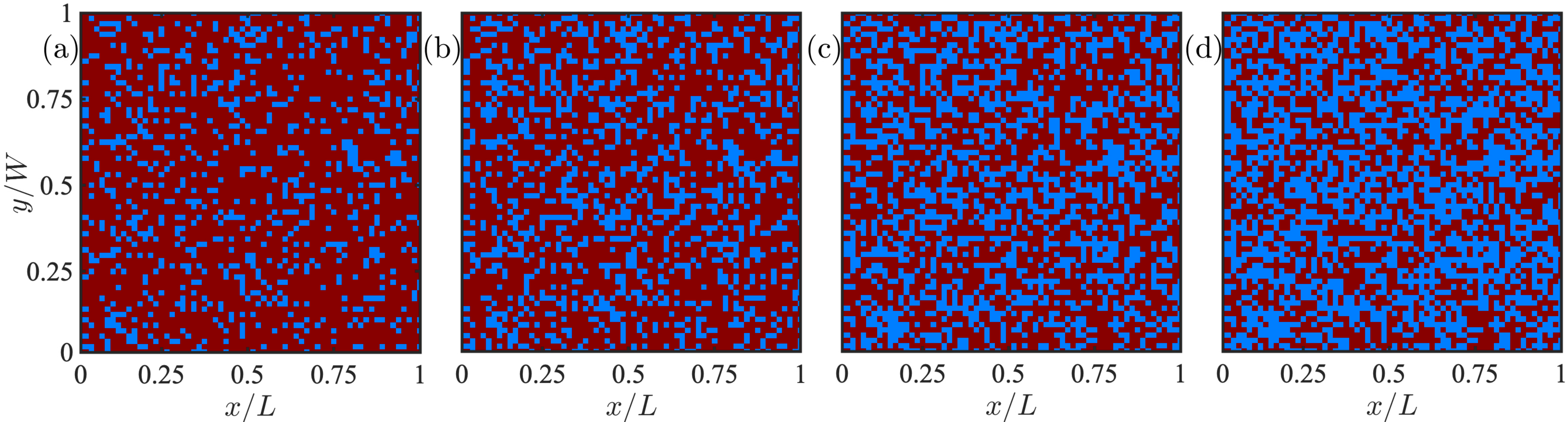}
\caption{Examples of the spatial distribution profile of random alloy systems in two dimensions $x$ and $y$. The red and blue colors represent the two different materials with different gap magnitudes in the band structure. In (a)-(d), the density of the compound made by the blue increases from $20\%$ to $50\%$ by a step of $10\%$.}
\label{samp}
\end{figure*}

\begin{figure*}
\centering
\includegraphics[width=\textwidth]{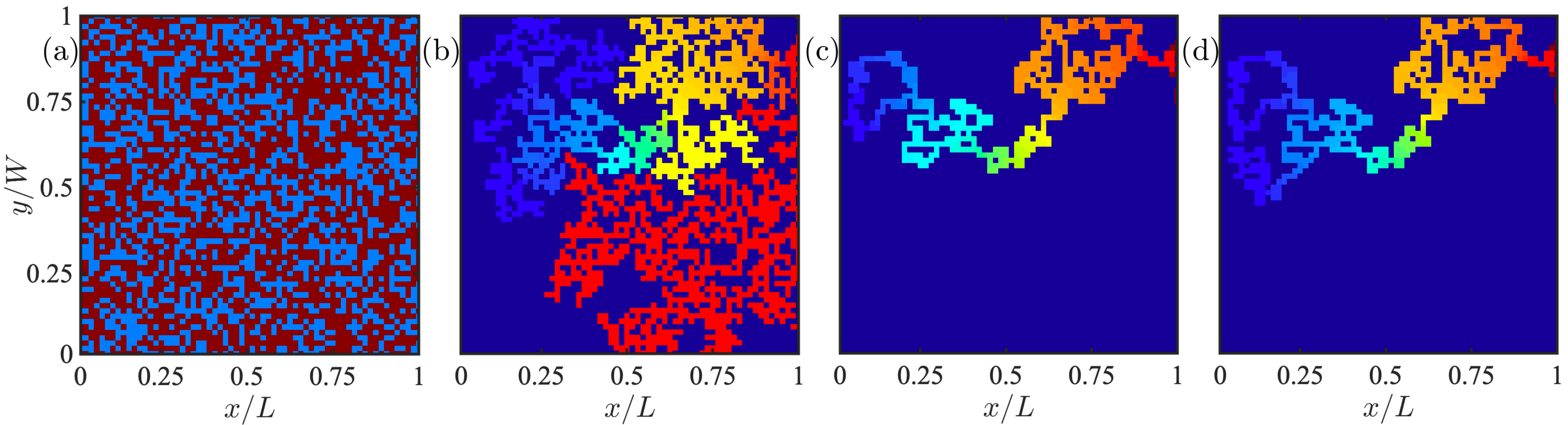}
\caption{The spatial map of self-consistent electrostatic potential $\varphi({\bm r})$. Panel (a) displays the random distribution of the two elements where the density of the blue regions (insulator or vacuum) is set fixed to $40\%$. Panel (b) exhibits $\varphi({\bm r})$ after two iterations of the self-consistent calculations whereas panel (c) shows the converged self-consistent $\varphi({\bm r})$. The electrostatic potential at the leftmost and rightmost edges along the $x$ direction is set at $V_1=0$ and $V_2=5$~eV, respectively, in (b) and (c). In panel (d), the same system is exposed to a larger voltage difference, i.e., $V_1=0$ and $V_2=15$~eV, and the converged self-consistent $\varphi({\bm r})$ is shown. In panels (b)-(d), the deep blue and deep red correspond to zero and the maximum value of potential, which is equal to $V_2$. }
\label{potential}
\end{figure*}

\subsection{The self-consistent algorithm}\label{algor}

The driving force for the moving charged particles across the sample is the electrostatic potential $\varphi({\bf r})$. However, in the presence of inhomogeneously distributed compositions with different thicknesses and bandgaps, the electrostatic potential naturally becomes spatially inhomogeneous and location dependent. Therefore, to accurately simulate a random alloy system, one needs to obtain the correct spatially inhomogeneous electrostatic potential $\varphi({\bf r})$. We have employed a self-consistent scheme to address this pivotal component of our approach. The flow chart of the operation of the self-consistent algorithm and the entire calculation approach are qualitatively summarized in Fig. \ref{flowchart}. When a random alloy system is created (e.g., shown in Fig.~\ref{samp}), we start with an initial guess to the spatial profile of the charge conductance. The final results and conclusions are independent of the initial guess to $g({\bf r})$. Therefore, we simply assign two constants $1$ and $0$ as initial guesses to the charge conductance of the two different materials, characterized by their bandgaps $\mu_2\gg\mu_1$ shown in Fig.~\ref{schem}. Starting with this initial guess, we solve Poisson's equation (\ref{eq:poisson}) and obtain $\varphi({\bf r})$. Next, accounting for the spatial distribution of the electrostatic potential $\varphi({\bf r})$, we solve the Schr\"{o}dinger equation (\ref{Miccurrent}) and obtain an updated map to the spatial charge conductance $g({\bf r})$. The process is iterated until the spatial profile of the charge conductance reaches stability and remains unchanged with a tolerance of $10^{-4}$ for at least 50 successive iterations. We repeat these self-consistent calculations for a few thousands of random samplings with similar densities of different materials, collect data, and take the average over the results to study the effective response of random alloys.

\subsection{Statistical sampling and generation of the two-phase random alloy} \label{sampling}

In order to create the random alloy systems, we have tested both (i) fully uncorrelated and (ii) correlated Monte Carlo samplings. In (i), the different sites of media belonging to a certain material are selected fully randomly, whereas in (ii), a natural correlation is employed to assign a certain material to a site. However, since we perform our calculations for few thousands of randomly generated samples, both sampling approaches result in identical results. Figure~\ref{samp} displays examples of the spatial profile of random alloy or porous systems. The system is assumed to be rectangular with length and width $L$ and $W$, respectively. The blue and red colors represent materials with different electronic characteristics (corresponding to different bandgaps in this paper). In Fig.~\ref{samp}(d), the density of the blue and red grains is equal: $50\%$. However, in Figs.~\ref{samp}(a), \ref{samp}(b), and \ref{samp}(c) the density of the blue grains increases  whereas the density of the red grains decreases: $20\%$($80\%$), $30\%$($70\%$), and $40\%$($60\%$), respectively.

\begin{figure*}
\centering
\includegraphics[width=\textwidth]{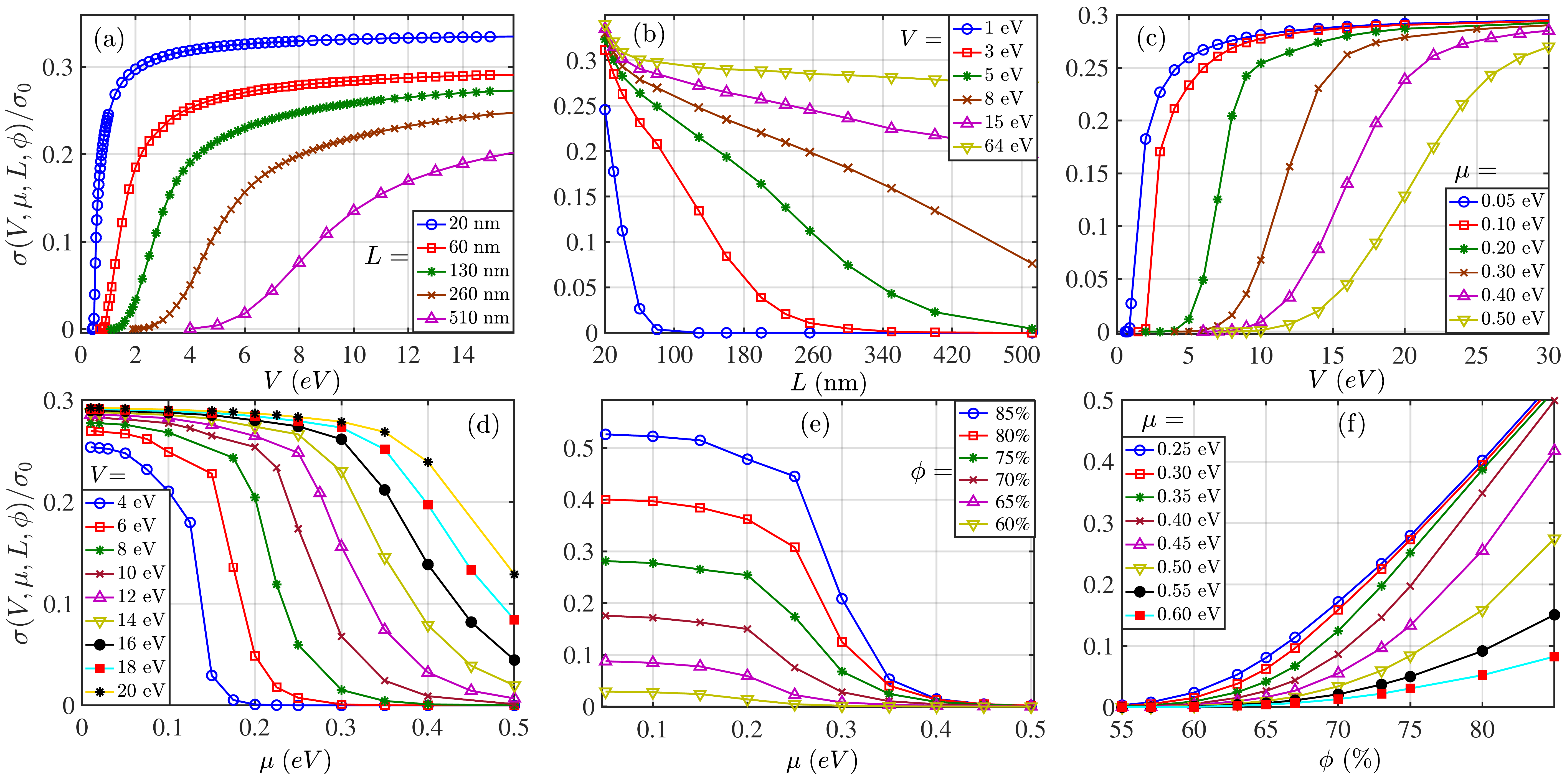}
\caption{Effective conductivity $\sigma (V,\mu,L,\phi)$. (a) The effective conductivity as a function of voltage difference $V$ where the system length and width are $W (\text{nm})=L (\text{nm})=20, 60, 130, 260, 510$. The bandgap and density of the moderate bandgap grains are set to $\mu=0.05$~eV and $\phi =75 \%$, respectively. (b) The effective conductivity as a function of junction length (W=L) for differing voltage differences $V=1,3,5,8,15,64$~eV and $\mu=0.05$~eV and $\phi=75\%$. (c) $\sigma$ as a function of voltage difference where $\mu=0.05,0.1,0.2,0.3,0.4,0.5$~eV, $L=64$~nm, and $\phi=75\%$. (d) The effective conductivity as a function of the bandgap $\mu$ for various voltage differences $V=4,6,8,10,12,14,16,18,20$~eV. The junction length is $L =64$~nm and the density is $\phi=75\%$. (e) $\sigma$ as a function of the bandgap where $\phi=85\% ,80\% ,75\% ,70\% ,65\% ,60\%$, $L=64$~nm, and $V=10$~eV. (f) The effective conductivity as a function of the density of the moderate bandgap grains $\phi$ where $\mu=0.25,0.3,0.35,0.4,0.45,0.5,0.55,0.6$~eV, $W=L=64$~nm, and $V=18$~eV. }
\label{cond}
\end{figure*}

\section{Results and Discussion}\label{results}
This section is divided into three subsections. In Sec.~\ref{sec:charge}, we present the results of our self-consistent approach for the charge conductance as a function of multiple parameters available in the system. Next, in Sec.~\ref{sec:func}, the parametric functionality of the effective charge conductance, averaged over numerous statistically produced samples, to the parameters of the system is given. Finally, in Sec.~\ref{sec:threshold}, through the analysis of our numerical results, the parametric functionality of threshold values, switching on and off the effective charge conductance, to the parameters of the system is obtained. 

\subsection{The effective self-consistent charge conductance}\label{sec:charge}
We start by illustrating the performance of the self-consistent approach we employ. Figure~\ref{potential} reveals the spatial profile of the electrostatic potential $\varphi({\bf r})$ in a representative random alloy or a porous system at different steps of the self-consistent iteration. Similar to Fig.~\ref{samp}, the vacancies (or insulator regions) with a large bandgap are blue whereas the semiconductor regions with a moderate bandgap are shown in red in Fig.~\ref{potential}(a). The density of semiconductor regions $\phi$ is considered to be $60\%$ ($40\%$). As displayed in Fig.~\ref{schem}(a), we consider a voltage difference $V=V_1-V_2$ between the two outer interfaces of the sample. However, as can be clearly seen in Fig.~\ref{potential}(a), there are only a few paths that connect continuously the left electrode to the right electrode through the red regions that allow for transporting charged particles through the electrostatic voltage difference from the left interface ($V_1$) to the right interface ($V_2$). The spatial profile of the electrostatic potential $\varphi({\bf r})$ after two iterations of the self-consistency loop is shown in Fig.~\ref{potential}(b). It is clearly seen that the spatial electrostatic potential starts to eliminate the unconnected regions and concentrate around percolated paths. After letting the self-consistent loop be iterated a hundred times, we find Fig.~\ref{potential}(c) for the spatial profile of $\varphi({\bf r})$. Remarkably, not only this approach successfully eliminates the unpercolated paths but it also finds the proper spatial distribution of the electrostatic potential. In the case of Figs.~\ref{potential}(b) and \ref{potential}(c), we have considered $V_1=0$ and $V_2=5$~eV. In Fig.~\ref{potential}, deep blue corresponds to zero and dark red is equal to $V_2$. The colorcode in the converged $\varphi({\bf r})$ in Fig.~\ref{potential}(c) reveals how the electrostatic potential drops from the right interface $V_2=5$~eV to the left interface $V_1=0$ when moving through the percolated path available in the system. Compared to Fig.~\ref{potential}(c), we have now increased the voltage difference to $V=15$~eV in Fig.~\ref{potential}(d). Note that for higher voltage differences, like what is shown in Fig.~\ref{potential}(d), more percolated paths are accessible due to stronger electron tunneling through the semiconductor regions. In the following, we consider square samples with equal sides, i.e., $W(\text{nm})=L(\text{nm})$.  
  
  \begin{figure*}
\centering
\includegraphics[width=\textwidth]{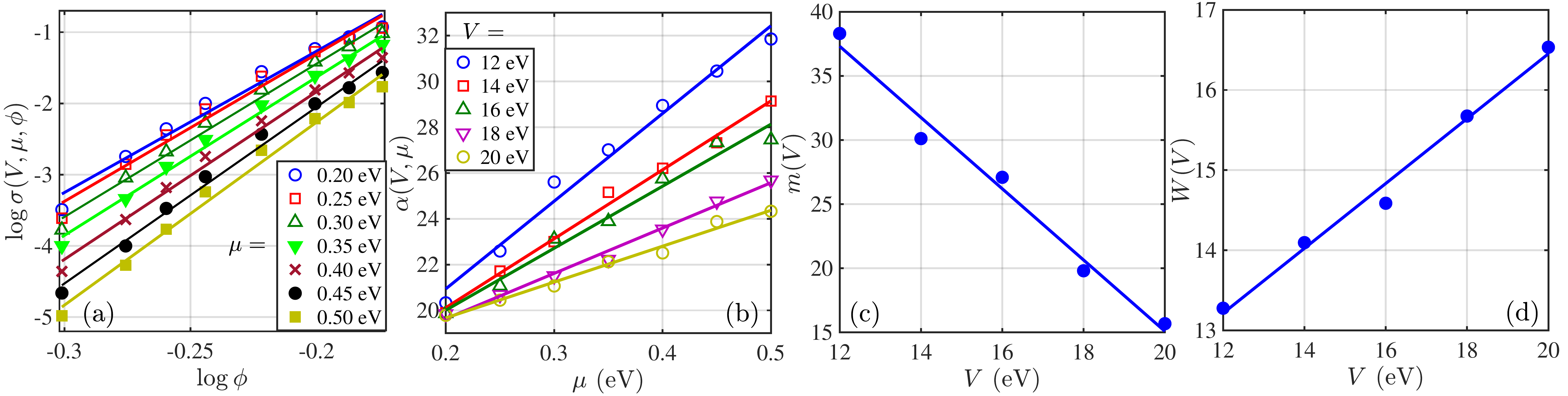}
\caption{(a) The log-log scale of conductivity against the density of the moderate bandgap grains $\phi$. (b) The slope of the log-log curves in panel (a) against the bandgap $\mu$ for different values of the voltage difference $V$. (c) and (d) are the slope and intersect of the curves shown in panel (b) as a function of the applied voltage difference.}
\label{sig_phi}
\end{figure*}

In Fig.~\ref{cond}, we study the effective charge conductance for all the experimentally accessible parameters available in our model, i.e., as a function of the voltage difference $V$, different junction area: $L(\text{nm})=20,60,130,260,510$, the bandgap $\mu$, and density $\phi$. The bandgap and density of semiconducting regions are $0.05$~eV and $\phi=75\%$, respectively, in Fig.~\ref{cond}(a). With increasing the voltage difference $V$, the charge conductance increases and reaches at a saturation value within high enough $V$ values. Figure~\ref{cond}(a) shows that the saturation voltage depends considerably on junction length. Figure~\ref{cond}(b) illustrates conductance as a function of $L(\text{nm})$ at various values of voltage difference. The charge conductance is suppressed exponentially with increasing the junction thickness and the suppression rate increases with decreasing the voltage difference $V$. This feature can be explained by noting the fact that the thickness of a semiconductor region prevents the tunneling of electrons and declines electron transmission. Additionally, we see that this approach allows for studying system size insulator regions of the order of a few thousand nanometers and depending on the other parameter values, the effective charge conductivity can be nonzero. In Fig.~\ref{cond}(c), the junction thickness is fixed at $L=64$~nm and the effective conductivity is plotted as a function of voltage difference $V$ for various values of the bandgap of the semiconductor regions $\mu$. As seen, increasing $\mu$ results in larger saturation voltages. To further illustrate the role of the bandgap, the effective conductivity as a function of $\mu$ for various voltage differences is plotted in Fig.~\ref{cond}(d). Increasing $\mu$, the charge conductivity drops exponentially and the suppression rate decreases with decreasing the voltage difference. The suppressing nature of the bandgap can be understood by considering the fact that the transmission of electrons through a barrier is suppressed by the height of the barrier: the higher the barrier, the lower the transmission probability. However, as the energy of particles increases, the transmission probability enhances and the particles acquire enough energy to tunnel through the quantum barrier.

In order to illustrate further features of the effective charge conductivity, we have plotted $\sigma$ as a function of the bandgap and the density of the moderate bandgap regions $\phi$ in Figs.~\ref{cond}(e) and \ref{cond}(f), respectively. The junction thickness is fixed at representative values $L=64$~nm and the voltage difference is set to $V=10$~eV and $V=18$~eV, respectively. Consistent with the previous findings, increasing the bandgap suppresses the effective conductivity and the increase of the density of the moderate bandgap regions enhances the effective charge conductivity. 

\begin{figure*}
\centering
\includegraphics[width=\textwidth]{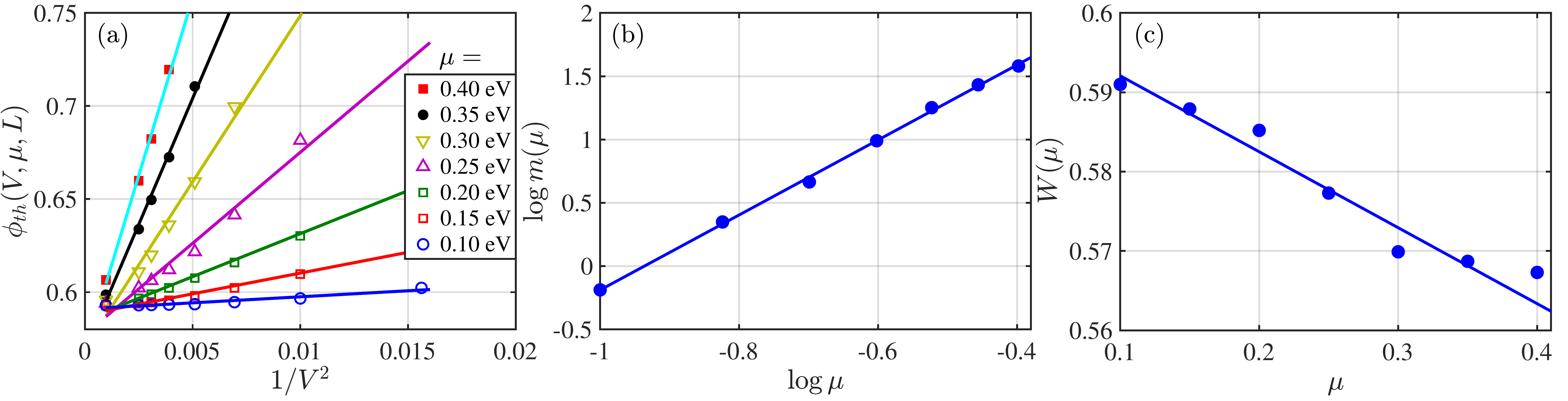}
\caption{(a) The threshold density of the moderate bandgap semiconductor grains as a function of voltage difference for various values of the bandgap value $\mu$. (b) and (c) The coefficients $\log m$ and $W$ used in fittings, shown in (a), as a function of the bandgap $\mu$. }
\label{phi_c}
\end{figure*}

\subsection{The functionality of the effective charge conductance to the parameters of the system}\label{sec:func}
Figure~\ref{cond} suggests that one is able to obtain simple relationships between the charge conductance and the independent parameters of the system, i.e., $V,\mu,L$, and $\phi$ within specific ranges of these parameters. Therefore, in what follows, we have restricted the material-vacancy density to $60\% \leq \phi \leq 85\%$, which is larger than the percolation threshold and less than the single phase material $\phi=100\%$. This range of material-vacancy density ensures nonzero finite quantum transport through the system and furthermore allows for obtaining simple power-law relations to the effective conductivity and threshold values as shall be presented below. Hence, we emphasize that the validity of the functionalities obtained below is limited to the mentioned specific ranges of parameter values and may be invalid otherwise. 

To determine these functionalities, we have plotted the charge conductivity as a function of the density of the moderate bandgap semiconducting regions in Fig.~\ref{sig_phi}(a). Keeping the length and width of the system constant at $L=64$nm simplifies our investigation. The log-log scale used suggests the following relationship:
\begin{equation}  
\log \sigma(V,\mu,\phi) = \alpha(V,\mu)\log \phi + \log \phi', \label{eq:sigphi}
\end{equation}  
in which $\alpha(V,\mu)$ is the slope of the log-log curves in Fig.~\ref{sig_phi}(a). To obtain the functionality, we have plotted  $\alpha(V,\mu)$ as a function of $\mu$ for various values of the voltage difference $V$ in Fig.~\ref{sig_phi}(b). The results suggest a linear dependence of $\alpha(V,\mu)$ to $\mu$ so that  
\begin{equation}  
\alpha(V,\mu) = m(V)\mu + W(V).
\end{equation}  
Incorporating these findings, we, therefore, end up with the following relationship for the charge conducting: 
\begin{subequations}
\begin{align}  
&\sigma(V,\mu,\phi) \propto \phi^{\alpha},\\
&\alpha(V,\mu) = \omega V\mu + \beta \mu + \zeta V + \gamma,
\end{align} 
\end{subequations}
 where $\omega =-2.78$, $\beta =70.65$, $\zeta =0.40$, and $\gamma =8.36$.

\subsection{The functionality of thresholds to the parameters of the system}\label{sec:threshold}

As it can be clearly seen in Fig.~\ref{cond}, all panels reveal threshold values to the experimentally relevant quantities $V,\mu,L,\phi$ where the effective conductivity acquires finite values otherwise is zero. By carefully studying these threshold values, one is able to obtain functionalities of these threshold values to $V,\mu,L,\phi$. In what follows, we perform such studies and obtain functionalities for two representative quantities, i.e., the threshold density and voltage difference.

To this end, we consider a system with fixed $L=64$~nm and plot the threshold density $\phi_{th}$ as a function of voltage difference for various values of the bandgap $\mu$. Our study found that $\phi_{th}$ is a linear function of $1/V^2$ as shown in Fig.~\ref{phi_c}(a). The dots show the calculated values for each case whereas the corresponding fitted linear dispersions are shown by solid lines. Therefore, Fig.\ref{phi_c}(a) suggests the following functionality for the threshold density to the voltage difference:
\begin{equation}
\phi_\text{th}(V,\mu) = \frac{m(\mu)}{V^2} + W(\mu),\label{eq:phi_c}
\end{equation} 
where $m(\mu)$ and $W(\mu)$ are the fitting parameters to be determined. To obtain the functionality of these two parameters to the bandgap parameter $\mu$, we have plotted the slope and intercept of the fitted linear dispersions of Fig.~\ref{phi_c}(a) in Figs.~\ref{phi_c}(b) and \ref{phi_c}(c) against the bandgap parameter $\mu$. Hence, we conclude the following functionality for $m$ and $W$ to $\mu$: 
\begin{equation}
\log m(\mu) = \alpha\log \mu + \log \mu', \label{eq:m}
\end{equation}
and 
\begin{equation}
W(\mu) = \gamma \mu +\eta. \label{eq:w}
\end{equation}
Submitting these functionalities Eqs.~(\ref{eq:m}) and (\ref{eq:w}) into Eq.~\ref{eq:phi_c} we obtain the functionality of the threshold density to the voltage difference and the bandgap as 
\begin{equation}
\phi_\text{th}(V,\mu) =\beta \frac{\mu^\alpha}{V^2} + \gamma \mu + \eta.
\end{equation}
Our calculations find precise values to the unknown parameters as $\log\beta=2.77$, $\alpha=2.96$, $\gamma=-0.089$, and $\eta=0.6$.

\begin{figure}[b]
\centering
\includegraphics[width=0.36\textwidth]{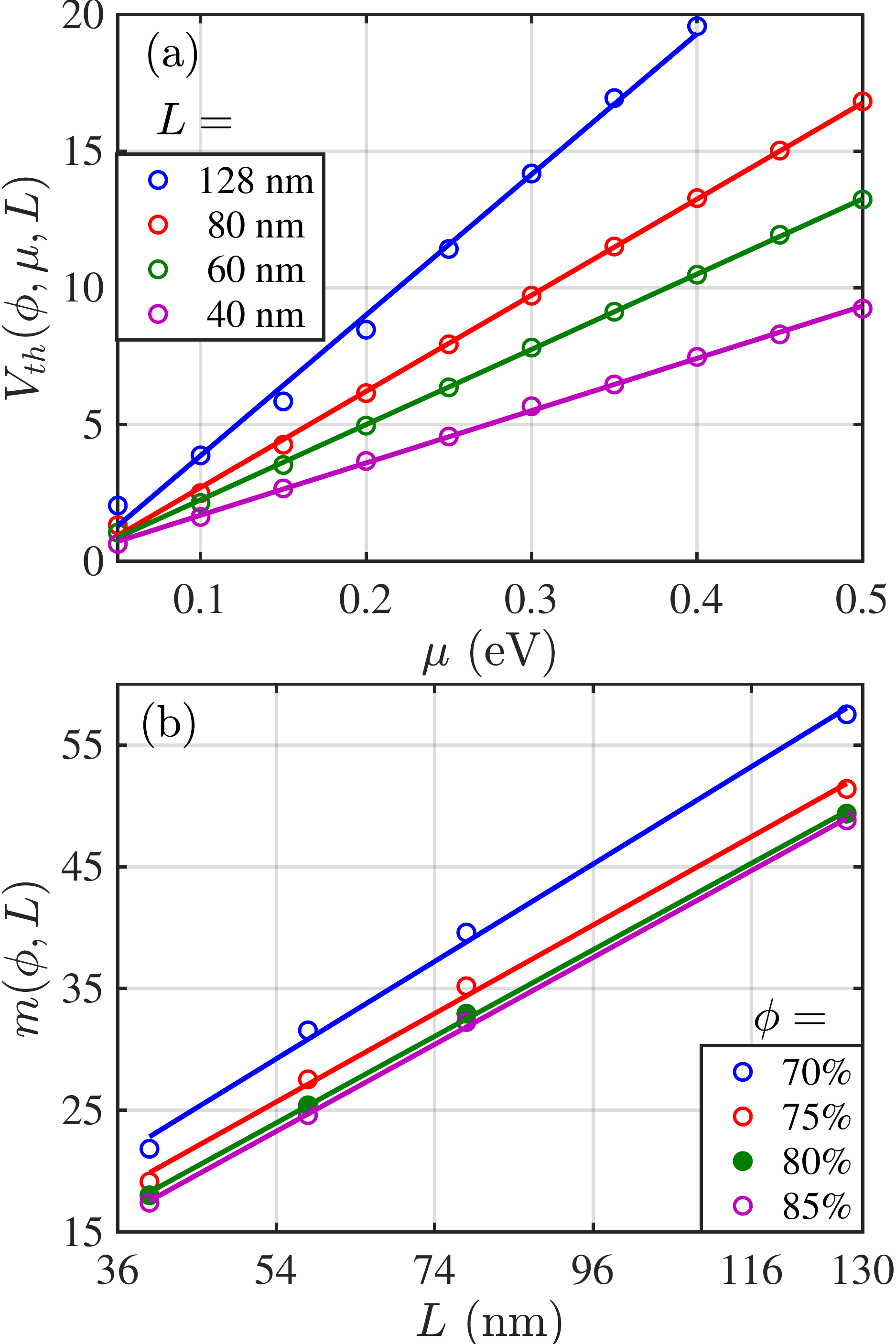}
\caption{(a) The threshold voltage difference against the bandgap $\mu$ for different values of junction thickness. (b) The slope of the linearly dispersing fitting shown in panel (a) as a function of junction thickness for differing values of the density of the moderate bandgap regions.}
\label{vc}
\end{figure}

Figures~\ref{cond}(a) and \ref{cond}(c) reveal threshold values to the voltage difference $V$ so that for voltages $V>V_\text{th}$, the charge conductance acquires finite values and is vanishingly small or zero when $V<V_\text{th}$. In our analysis and determining $V_\text{th}$, the charge conductance is considered finite where the statistical averaging of $\sigma$ is less than $0.01$.  
 As seen in Fig.~\ref{cond}, $V_\text{th}$ is a function of $\phi, \mu$, and $L$. In order to determine this functionality, we have plotted the threshold voltage difference $V_\text{th}$ vs the bandgap $\mu$ for $L=40,60,80,120$~nm in Fig.~\ref{vc}(a). The density of the moderate bandgap regions is set fixed to $75\%$. The calculated values through the self-consistent algorithm are marked by dots whereas the fitted dispersions are shown by solid lines. As clearly seen, the results suggest a linear functionality to $\mu$ as follows: 
\begin{equation}
V_\text{th}(\phi,\mu,L) \propto m(\phi,L) \mu,
\end{equation}
where $m$ is the slope of the fitted linear dispersions dependent on $\phi,L$. To further determine the functionality of this parameter, we have plotted it as a function of $L$. Figure~\ref{vc}(b) displays $m$ against $L$ for differing values of the density of the moderate bandgap regions $\phi$. The results reveal a linear dispersion as follows:
\begin{equation}
m(\phi,L) =a(\phi) L + L'.
\end{equation}
Obtaining the slope of $m(\phi,L)$ as a function of $\phi$, we determine its functionality to the density of the moderate bandgap regions. After performing the fitting process, we find the following functionality for the threshold voltage to $\mu, L, \phi$,  
\begin{equation}
V_\text{th}( \phi,\mu,L) \propto  (\alpha \phi + \beta)^4\mu L + \gamma\mu L, 
\end{equation}
in which $\alpha=3.404$, $\beta=-2.84$, and $\gamma=0.356$. 

This parametric study can be easily reperformed for realistic systems and the threshold values and their functionality to the considered parameters be tested to confirm the correctness of the predictions made in this paper. Also, the expansion of the parametric study to obtain more generalized functionalities to the threshold values will be the topic of a future work. 

\section{conclusions}\label{conclusion}

We have developed a multiscale self-consistent algorithm, incorporating quantum mechanical charge transport and Monte Carlo sampling, to study the charge transport through a random alloy with vacancies or porous systems. Starting from an initial guess to the spatial dependency of the local conductivity of random grains quantum mechanically $g(\textbf{r})$, the algorithm calculates spatial driven electrostatic voltage $\varphi(\textbf{r})$, computes a new $g(\textbf{r})$ throughout the sample by the charge conservation law, and repeats these two steps until the error in calculating $g(\textbf{r})$ reaches a tolerance value. Employing this algorithm, we study the effective charge conductance parametrically in a random alloy or porous system made of two different phases: (i) moderate bandgap compound and (ii) vacancy (insulator). Our results reveal that by accounting for the quantum mechanical aspects of the system, a highly nonlinear charge conductance as a function of voltage difference $V$, junction thickness $L$, the density of moderate bandgap compound $\phi$, and the bandgap of, e.g., intermetallic compound phase $\mu$ appear. Our findings of the effective charge conductance show that the heterogeneous system passes no current within specific regions of parameter space and is finally activated for threshold values of $V, L, \phi,$ and $\mu$. We find that the threshold value of one quantity $V, L, \phi,$ and $\mu$ strongly depends on the other quantities. Through an exhaustive computational study, as representative cases, we determine the functionality of the threshold density of the moderate bandgap compound $\phi_\text{th}$ to $V$ and $\mu$ and also the threshold of voltage difference $V_\text{th}$ to $\phi, L, \mu$. 

The results of our parametric study can be employed as guidelines to design future electronic devices containing elements that can be described by a random porous scenario or alloy systems with vacancies.
Once, in an experiment, the details mainly on the exact values of $\mu$, $\phi$, $V$, and $L$ are available, one can simply substitute them into the obtained simple functionalities, calibrate the formulas, and confirm their validity.

\section*{DATA AVAILABILITY}

The data that support the findings of this study are available from the authors upon reasonable request.

\section*{Author contributions}

All the authors contributed to the project. E.S. performed the main calculations in discussion with M.A. and H. H. and prepared the figures. All the authors contributed to analyzing the results and data. M.A. prepared the first draft of the results' text and all the authors reviewed and edited the entire manuscript text.

\end{document}